\documentclass[letterpaper,11pt]{article}%
\usepackage{amsmath}
\usepackage{epsfig}
\usepackage{graphicx}%
\usepackage{amsfonts}%
\usepackage{amssymb}

\begin{document}

\title{{\Huge A New Basis for QED }\\{\Huge Bound State Computations}\\\ \\\ \ \\\ }
\author{J. Pestieau and C. Smith
\and \ \ \ \ \ \ \ \medskip\\\textsl{Institut de Physique Th\'{e}orique, Universit\'{e} catholique de
Louvain,}\linebreak \\\ \ \textsl{\linebreak Chemin du Cyclotron, 2, B-1348, Louvain-la-Neuve}\\\ \ }
\date{November 28, 2001}
\maketitle

\begin{abstract}
A simple method to compute QED bound state properties is presented, in which
binding energy effects are treated non-perturbatively. It is shown that to
take the effects of all ladder Coulomb photon exchanges into account, one can
simply perform the derivative of standard QED amplitudes with respect to the
external momentum. For example, the derivative of the light-by-light
scattering amplitude gives an amplitude for orthopositronium decay to three
photons where any number of Coulomb photon exchanges between the $e^{+}e^{-}$
is included.

Various applications are presented. From them, it is shown that binding energy
must be treated non-perturbatively in order to preserve the analyticity of
positronium decay amplitudes.

Interesting perspectives for quarkonium physics are briefly sketched.

\pagebreak 

\end{abstract}

\section{Introduction}

The properties of positronium, the bound state made of an electron and a
positron, provide some of the most precise tests of QED \cite{QEDTest}. Both
the experimental \cite{Experm} and theoretical considerations have reach a
very high level of precision, requiring for the latter the computations of
many-loop diagrams (\cite{RadiativeCorr}-\cite{Anton}, and references cited there).

In the present note, we will address one particular aspect of the current QED
bound state models, namely the factorization between the bound state dynamics
and the decay processes \cite{PreviousWork}. As explained in
\cite{PreviousWork2}, the present study is motivated by the recurrent
contradiction between standard positronium models and Low's theorem
\cite{Low}. Our goal is to show that binding energy effects cannot be treated
perturbatively if sensible theoretical results are sought at the present level
of precision. The central result of the present note is a simple method that
allows such an exact non-perturbative treatment of binding energy effects. The
lowest order results we shall obtain sum infinite class of diagrams, usually
expanded perturbatively. Also, our method is formally relativistic and has a
correct analytical behavior.

In section $2$, our simple method is introduced by a computation of the rate
for parapositronium to two photons, and then demonstrated in general. In the
next sections, we apply our result to various decay modes chosen to illustrate
some specific aspects.

First, to show explicitly the contradiction between Low's theorem and binding
energy perturbative expansions, we consider the paradimuonium ($p$-$Dm$, a
bound $\mu^{+}\mu^{-}$ \cite{Dimuonium}) decay to $\gamma e^{+}e^{-} $. Then,
interesting connections between bound state decay amplitudes and the photon
vacuum polarization are exemplified by the orthodimuonium decay $o$%
-$Dm\rightarrow e^{+}e^{-}$.

The fourth section is devoted to the interesting decay mode orthopositronium
to three photons. We obtain the differential spectrum and decay rate for any
value of the binding energy, from the amplitude for light-by-light scattering.
We discover that both the spectrum and rate are highly sensitive to the
binding energy when it is approaching zero. The implication of this result is
the well-known slow convergence of the orthopositronium radiative correction
perturbative series.

In the fifth section, we present two extensions of our method. The application
to orthopositronium $n$th radial excitations $o$-$Ps\left(  nS\right)  $ is
introduced by recomputing the decay rate $\pi^{0}\rightarrow\gamma
\,o$-$Ps\left(  nS\right)  $ \cite{Pion}. Finally, it is shown how the
positronium hyperfine splitting, i.e. the mass difference of the ortho- and
parapositronium, can be obtained from vacuum polarization graphs through
finite mass renormalizations.

\section{Lowest Order Decay Amplitudes}

Positronium amplitudes are built as loop amplitudes: the positronium couples
to a virtual $e^{+}e^{-}$ loop, to which a given number of photons are
attached. The coupling of the positronium to its constituents is essentially
determined by the positronium wavefunction. In the following, we will take a
form factor inspired from the Schr\"{o}dinger wavefunction, i.e. a form factor
containing the effects of the exchange of Coulomb photons among the
constituents (in the ladder approximation) \cite{BarbRem}.

A very simple method can be used to compute such loop amplitudes: \textit{any
lowest order loop amplitude with a Coulomb form factor is the derivative with
respect to the positronium mass of the corresponding point-like amplitude}. By
point-like amplitude is meant the loop amplitude obtained by replacing the
complicated Coulomb form factor for the bound state by a constant from factor.
Symbolically, for parapositronium (orthopositronium) decay to an even (odd)
number of real or virtual photons, the amplitude is
\[%
\raisebox{-0.5154in}{\includegraphics[
height=1.0914in,
width=1.5506in
]%
{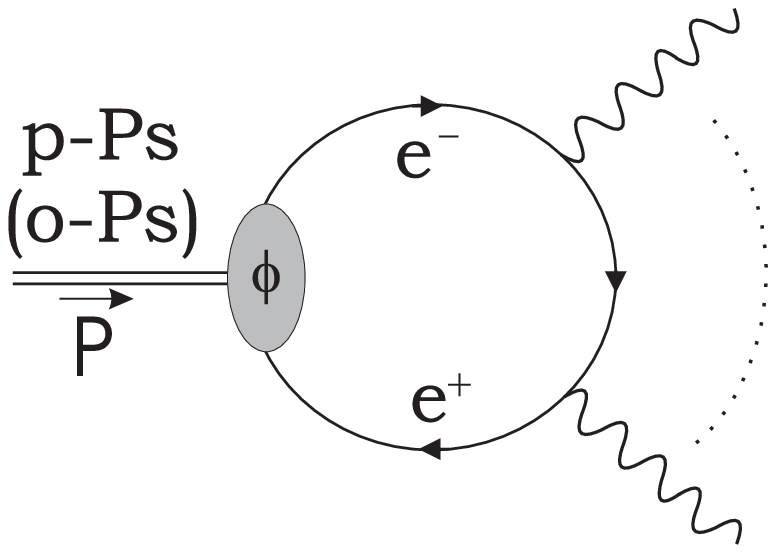}%
}%
\propto\frac{\partial}{\partial P^{2}}\left[
\raisebox{-0.4999in}{\includegraphics[
height=1.094in,
width=1.5506in
]%
{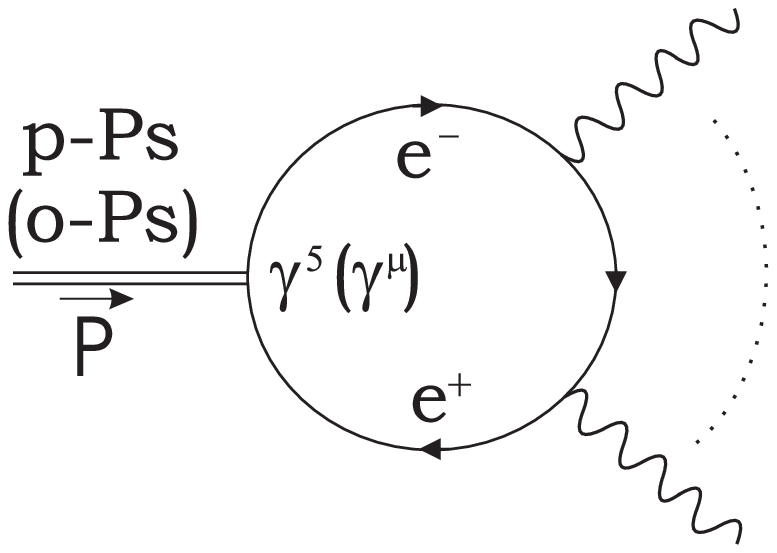}%
}%
\right]
\]
where $\phi$ represents the Schr\"{o}dinger wavefunction, i.e.
\[%
\raisebox{-0.5076in}{\includegraphics[
height=1.0914in,
width=1.5584in
]%
{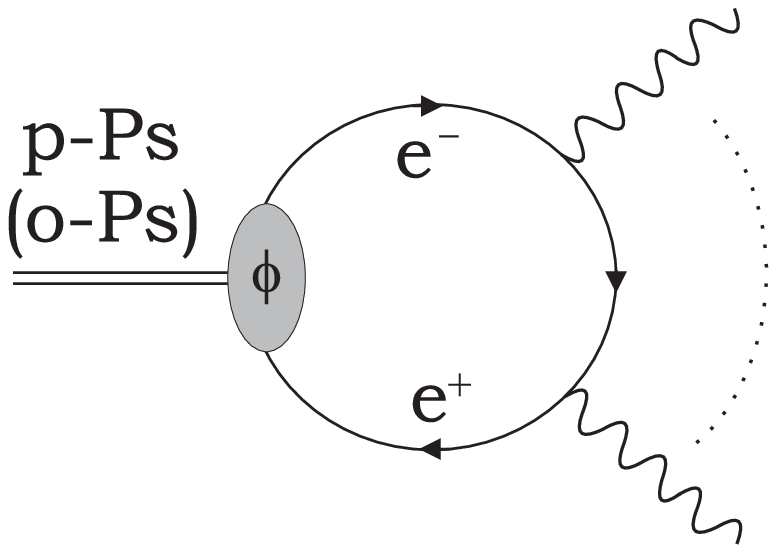}%
}%
=%
\raisebox{-0.4999in}{\includegraphics[
height=1.0905in,
width=1.5558in
]%
{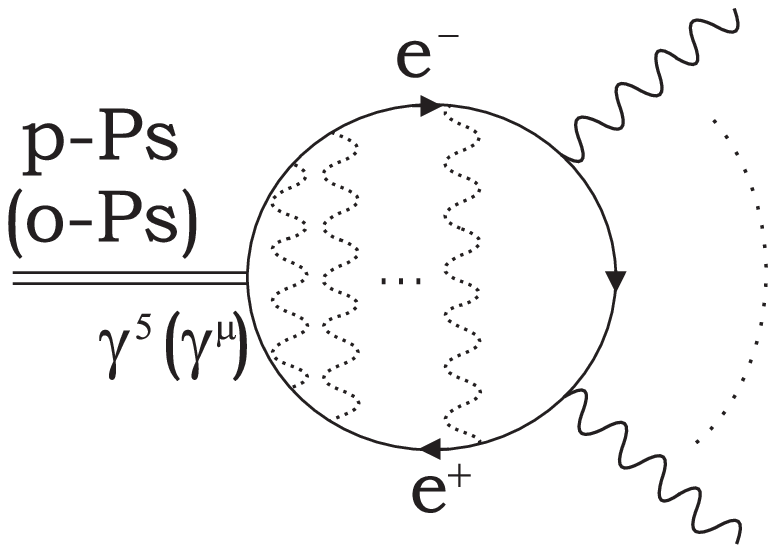}%
}%
\]
and dashed photon lines stand for Coulomb photons.

In the next subsection, we give an example to illustrate explicitly how the
method works, and in the following one, we present the general demonstration.

\subsection{An Example: Parapositronium to two photons}

The point-like amplitude for the decay $p$-$Ps\rightarrow\gamma\gamma$ is
obtained by replacing the Coulomb form factor by an elementary, point
particle, with a pseudoscalar $\gamma_{5}$ coupling to the electron current
(with unit coupling constant)%
\[%
{\includegraphics[
height=1.1156in,
width=1.574in
]%
{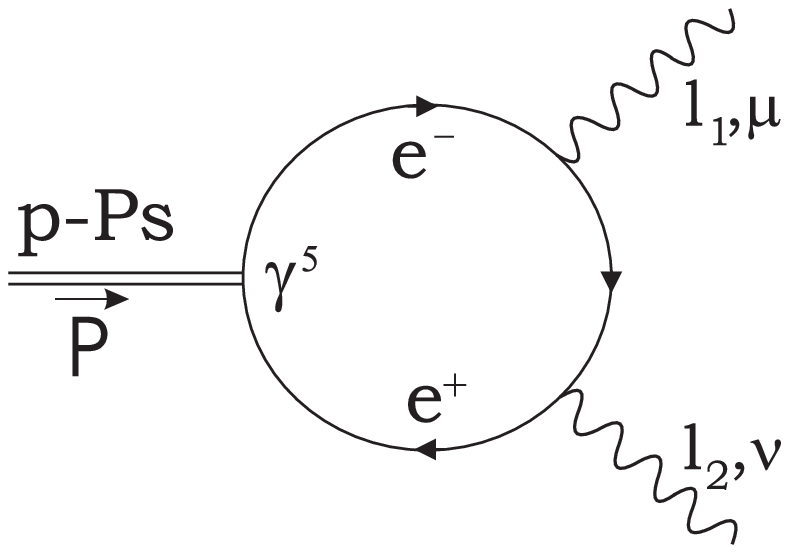}%
}%
\]
plus the crossed process. The point bound state amplitude is fully
relativistic, and standard loop integration techniques can be used.
Explicitly, the point-like amplitude is (the subscript $\gamma_{5}$ is a
reminder for point-like):
\begin{align*}
\mathcal{M}_{\gamma_{5}}\left(  p\text{-}Ps\rightarrow\gamma\gamma\right)   &
=ie^{2}\int\frac{d^{4}q}{\left(  2\pi\right)  ^{4}}Tr\left\{  \Gamma^{\mu\nu
}\right\}  \varepsilon_{\mu}\left(  l_{1}\right)  \varepsilon_{\nu}\left(
l_{2}\right) \\
\Gamma^{\mu\nu}  &  =\gamma_{5}\frac{1}{%
\!\not\!%
q+%
\!\not\!%
l_{1}-m}\gamma_{\mu}\frac{1}{%
\!\not\!%
q-m}\gamma_{\nu}\frac{1}{%
\!\not\!%
q-%
\!\not\!%
l_{2}-m}\\
&  +\gamma_{5}\frac{1}{%
\!\not\!%
q+%
\!\not\!%
l_{2}-m}\gamma_{\nu}\frac{1}{%
\!\not\!%
q-m}\gamma_{\mu}\frac{1}{%
\!\not\!%
q-%
\!\not\!%
l_{1}-m}%
\end{align*}
Carrying the trace, we readily obtain%
\[
\mathcal{M}_{\gamma_{5}}\left(  p\text{-}Ps\rightarrow\gamma\gamma\right)
=-8me^{2}\varepsilon^{\mu\nu\rho\sigma}l_{1,\rho}l_{2,\sigma}\varepsilon_{\mu
}\left(  l_{1}\right)  \varepsilon_{\nu}\left(  l_{2}\right)
\frac{\mathcal{I}_{\gamma_{5}}\left(  M^{2}\right)  }{M^{2}}%
\]
where the \textbf{dimensionless} loop integral form factor is
\begin{equation}
\mathcal{I}_{\gamma_{5}}\left(  P^{2}\right)  =\frac{-i}{\left(  4\pi\right)
^{2}}F\left[  \frac{4m^{2}}{P^{2}}\right]  \text{ \ \ \ with\ }F\left[
a\right]  =2\arctan^{2}\left(  a-1\right)  ^{-1/2} \label{LoopFF}%
\end{equation}
From the amplitude, the decay width is%
\begin{equation}
\Gamma_{\gamma_{5}}\left(  p\text{-}Ps\rightarrow\gamma\gamma\right)
=16\pi\alpha^{2}\,\frac{m^{2}}{M}\,\left|  \mathcal{I}_{\gamma_{5}}\left(
M^{2}\right)  \right|  ^{2}=\dfrac{\pi M\alpha^{2}}{256}\left(  \frac{4\gamma
^{2}}{M^{2}}+1\right)  \,\left(  \dfrac{4}{\pi^{2}}\arctan^{2}\dfrac
{M}{2\gamma}\right)  ^{2} \label{RatePunct}%
\end{equation}
where $\gamma^{2}=m^{2}-M^{2}/4$ is related to the binding energy $E_{B}=M-2m$.

To get the physical positronium decay amplitude and rate, simply replace in
(\ref{RatePunct}) the loop form factor $\mathcal{I}_{\gamma_{5}}\left(
M^{2}\right)  $ by its derivative%
\begin{equation}
\fbox{$\mathcal{I}_{Coul}\left(  M^{2}\right)  =\left(  32\pi C\phi_{o}%
\gamma\right)  \dfrac{\partial}{\partial M^{2}}\mathcal{I}_{\gamma_{5}}\left(
M^{2}\right)  $} \label{Derivative}%
\end{equation}
where $\phi_{o}$ is the $S$-wave fundamental state Schr\"{o}dinger
wavefunction at zero separation, and $C=\sqrt{M}/m$ is obtained by matching
the static limit $\left(  \gamma\rightarrow0\right)  $ with the well-known
lowest order result $\Gamma_{p\text{-}Ps}=m\alpha^{5}/2$ \cite{PreviousWork}.
This gives%
\[
\mathcal{I}_{Coul}\left(  M^{2}\right)  =-i\frac{C\phi_{o}}{M}\left[
\frac{2}{\pi}\arctan\frac{M}{2\gamma}\right]
\]
The factor in square brackets is equal to $1$ in the limit $\gamma
\rightarrow0$. Using $\left|  \phi_{o}\right|  ^{2}=\alpha^{3}m^{3}/8\pi$, the
decay rate into two-photon is%
\begin{equation}
\Gamma\left(  p\text{-}Ps\rightarrow\gamma\gamma\right)  =\frac{\alpha^{5}%
m}{2}\,\left(  \frac{4m^{2}}{M^{2}}\right)  \,\left|  \frac{2}{\pi}%
\arctan\frac{M}{2\gamma}\right|  ^{2} \label{TwoPhot}%
\end{equation}
The result (\ref{TwoPhot}) contains the effects of Coulomb interactions among
the constituents, at all orders in $\alpha$. Indeed, using $\gamma^{2}%
=m^{2}-M^{2}/4\approx m^{2}\alpha^{2}/4$, the form factor can be expanded as
(see the discussion in \cite{PreviousWork})
\[
\Gamma\left(  p\text{-}Ps\rightarrow\gamma\gamma\right)  =\frac{\alpha^{5}%
m}{2}\left(  1-2\frac{\alpha}{\pi}+\mathcal{O}\left(  \alpha^{2}\right)
\right)  \approx\frac{\alpha^{5}m}{2}\left(  1-0.637\alpha+\mathcal{O}\left(
\alpha^{2}\right)  \right)
\]
In other words, the binding energy effects included in our lowest order
computation already account for a great deal of the relativistic and radiative
corrections as presented in the literature \cite{RadiativeCorr}:%
\begin{equation}
\Gamma_{p\text{-}Ps}=\frac{\alpha^{5}m}{2}\left(  1-\delta\Gamma_{p\text{-}%
Ps}\right)  \label{ParaCorr}%
\end{equation}
where
\begin{align}
\delta\Gamma_{p\text{-}Ps}  &  =A_{p}\frac{\alpha}{\pi}+2\alpha^{2}\ln
\frac{1}{\alpha}+B_{p}\frac{\alpha^{2}}{\pi^{2}}-\frac{3\alpha^{3}}{2\pi}%
\ln^{2}\frac{1}{\alpha}+C_{p}\frac{\alpha^{3}}{\pi}\ln\frac{1}{\alpha}%
+\delta_{4\gamma}\frac{\alpha^{2}}{\pi^{2}}\label{ParaSeries}\\
&  \text{with \ \ }%
\begin{tabular}
[c]{l}%
$A_{p}=5-\pi^{2}/4\approx2.5326$\\
$B_{p}=5.14\left(  30\right)  $%
\end{tabular}
\ \;\;%
\begin{tabular}
[c]{l}%
$C_{p}=-7.919\left(  1\right)  $\\
$\delta_{4\gamma}=0.274\left(  1\right)  $%
\end{tabular}
\nonumber
\end{align}
Numerically, we can write (\ref{ParaCorr}) as%
\[
\Gamma_{p\text{-}Ps}\approx\frac{\alpha^{5}m}{2}\frac{4m^{2}}{M^{2}}\left(
\frac{2}{\pi}\arctan\frac{M}{2\gamma}\right)  ^{2}\left(  1-\delta
\Gamma_{p\text{-}Ps}^{\prime}\right)
\]
with the same series \ref{ParaSeries}, but with the reduced coefficients
\[%
\begin{tabular}
[c]{l}%
$A_{p}^{\prime}\approx0.5326$\\
$B_{p}^{\prime}\approx0.607$\\
$C_{p}^{\prime}\approx-3.919$%
\end{tabular}
\]
This last form is very interesting because binding energy corrections (i.e.
$\gamma$-dependent) are singled out, while the remaining radiative corrections
are much reduced. This means that one could, at least in principle, express
the decay rate as non-perturbative binding energy corrections times a rapidly
converging perturbation series of radiative corrections. This is exactly what
we have achieved to order $\alpha$.

\subsection{Generalization}

The proof of (\ref{Derivative}) is straightforward using the language of
dispersion relations. Indeed, the point-like loop amplitude can be computed
from its imaginary part
\[
\operatorname{Im}\mathcal{T}_{\gamma_{5}}\left(  P^{2}\right)  \equiv
\operatorname{Im}\mathcal{M}_{\gamma_{5}}\left(  p\text{-}Ps\left(
P^{2}\right)  \rightarrow\gamma\gamma\right)
\]
using an unsubstracted dispersion relation with $s=P^{2}$:
\begin{equation}
\mathcal{T}_{\gamma_{5}}\left(  M^{2}\right)  =\operatorname{Re}%
\mathcal{T}_{\gamma_{5}}\left(  M^{2}\right)  =\frac{1}{\pi}\int_{4m^{2}%
}^{\infty}\frac{ds}{s-M^{2}}\operatorname{Im}\mathcal{T}_{\gamma_{5}}\left(
s\right)  \label{DispInt}%
\end{equation}
($\mathcal{T}\left(  M^{2}\right)  =\operatorname{Re}\mathcal{T}\left(
M^{2}\right)  $ because $M<2m$). The Schr\"{o}dinger form factor accounting
for the non-trivial coupling of the bound state to its constituent is of the
form
\[
F_{B}\left(  0,\mathbf{q}\right)  =C\phi_{o}\mathcal{F}\left(  \mathbf{q}%
^{2}\right)  \left(  \mathbf{q}^{2}+\gamma^{2}\right)  \text{
\ \ \ with\ \ \ }\mathcal{F}\left(  \mathbf{q}^{2}\right)  =\frac{8\pi\gamma
}{\left(  \mathbf{q}^{2}+\gamma^{2}\right)  ^{2}}%
\]
where one can recognize $\phi\left(  \mathbf{q}^{2}\right)  \equiv\phi
_{o}\mathcal{F}\left(  \mathbf{q}^{2}\right)  $ as the fundamental $(n=1)$
$S$-wave Schr\"{o}dinger momentum space wavefunction. When expressed in terms
of the dispersion relation variable, this form factor is only a function of
$s$, the initial energy:%
\[
F_{B}\left(  s\right)  =C\phi_{o}\frac{32\pi\gamma}{s-M^{2}}%
\]
The core of the derivative approach emerges from the observation that
inserting $F_{B}$ in (\ref{DispInt}) is equivalent to taking the derivative
with respect to $M^{2}$
\begin{align}
\mathcal{T}_{Coul}\left(  M^{2}\right)   &  =\frac{1}{\pi}\int_{4m^{2}%
}^{+\infty}\frac{ds}{s-M^{2}}F_{B}\left(  s\right)  \operatorname{Im}%
\mathcal{T}_{\gamma_{5}}\nonumber\\
&  =\left(  32\pi C\phi_{o}\gamma\right)  \frac{1}{\pi}\int_{4m^{2}}^{+\infty
}\frac{ds}{\left(  s-M^{2}\right)  ^{2}}\operatorname{Im}\mathcal{T}%
_{\gamma_{5}}\nonumber\\
&  =\left(  32\pi C\phi_{o}\gamma\right)  \frac{\partial}{\partial M^{2}%
}\mathcal{T}_{\gamma_{5}}\left(  M^{2}\right)  \label{Derivative2}%
\end{align}
which is the desired result. The case of other parapositronium decay channels,
or orthopositronium decay modes is similarly treated (simply replace
$\gamma^{5}$ by $%
\!\not\!\!\!\
e$ with $e^{\mu}$ the orthopositronium polarization vector).

As a first remark, we repeat here the conclusion of \cite{PreviousWork}, which
is that standard positronium result are recovered from the above dispersion
integral (\ref{DispInt}) provided only the vertical cuts are taken into
account in the imaginary part.\ This proves that some contributions are missed
in those computations. For instance, in the case of $p$-$Ps\rightarrow
\gamma\gamma$, there is only the vertical cut, but this is not the case in
general: if one of the photon is virtual, or if there is three or more photons
in the final state, oblique cuts contribute (see next sections).

The second remark concerns the insertion of the form factor, not in Feynman
amplitudes, but directly into the dispersion integral. These two approaches
are equivalent only in the punctual case (consider the momentum flow through
the form factor in each case). Working at the level of dispersion relations is
much more in the spirit of the Bethe-Salpeter equation. Indeed, the
wavefunction is extracted from the four-point Green's function, i.e. in
configurations with an off-shell bound state (above threshold), and on-shell
constituents:%
\[%
{\includegraphics[
height=0.6711in,
width=1.9363in
]%
{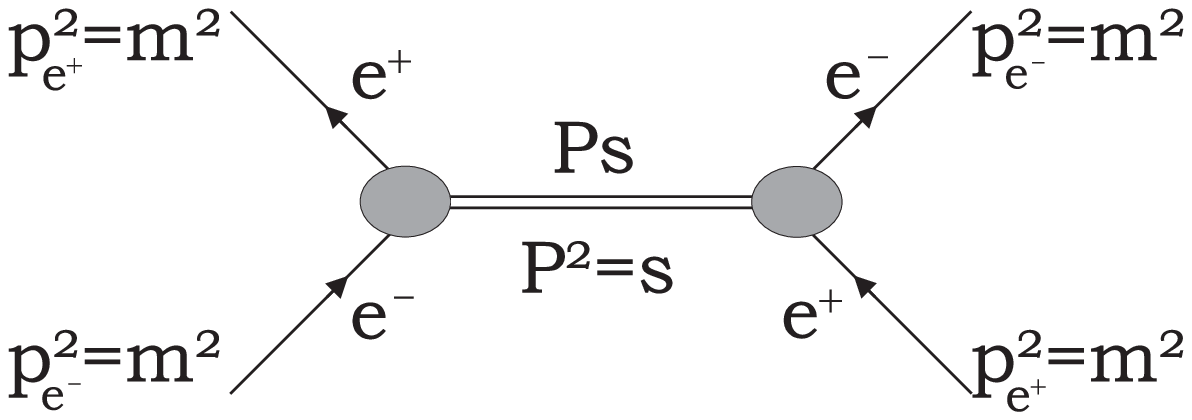}%
}%
\]
The picture shows that both Bethe-Salpeter and the dispersion integral
(\ref{DispInt}) make use of the form factor $F_{B}$ with the same kinematical
configuration (diagrams with bremsstrahlung radiation off the electron lines
are treated similarly).

\section{Application to decay rate computations}

We will now review some applications of the result (\ref{Derivative2}). We
leave the interesting case of $o$-$Ps\rightarrow\gamma\gamma\gamma$ to the
next section.

\subsection{Paradimuonium and Low's Theorem}

The paradimuonium decay $p$-$Dm\rightarrow\gamma e^{+}e^{-}$ is the simplest
QED bound state decay process where Low's theorem implications can be
illustrated \cite{PreviousWork}, \cite{PreviousWork2}, \cite{Low}. Following
the same steps as for $p$-$Ps\rightarrow\gamma\gamma$, we first consider the
point-like amplitude%
\[%
{\includegraphics[
height=1.1761in,
width=2.0141in
]%
{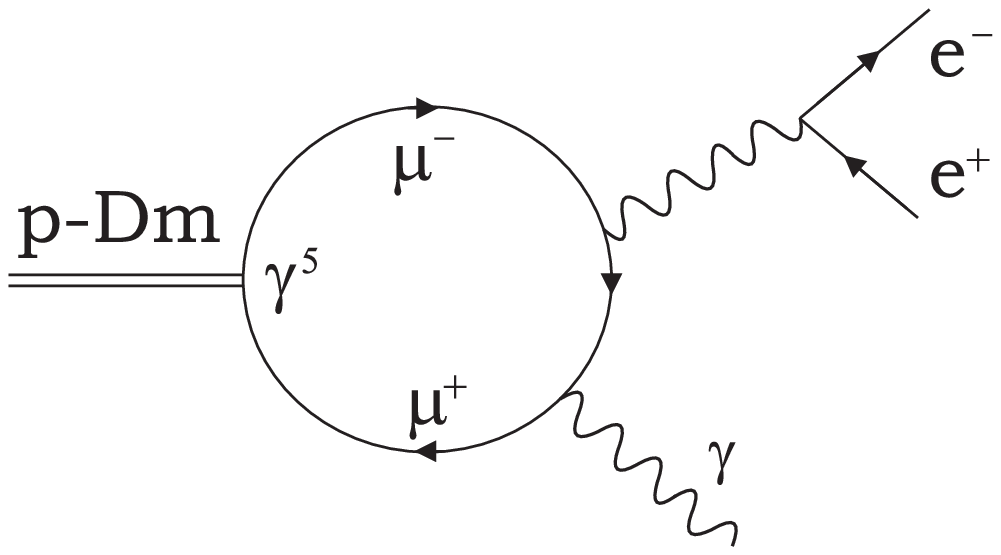}%
}%
\]
(plus the crossed diagram) with the result
\begin{align}
\Gamma_{\gamma_{5}}\left(  p\text{-}Dm\rightarrow e^{+}e^{-}\gamma\right)   &
=\dfrac{16\alpha^{3}}{3}\frac{m^{2}}{M}%
{\displaystyle\int\limits_{0}^{1-a_{e}}}
dx_{\gamma}\left|  \mathcal{I}_{\gamma_{5}}\left(  M^{2},x_{\gamma}\right)
\right|  ^{2}\rho\left(  x_{\gamma},a_{e}\right) \label{DalitzPhoton}\\
\text{with \ \ \ \ }\rho\left(  x_{\gamma},a_{e}\right)   &  =\sqrt
{1-\frac{a_{e}}{1-x_{\gamma}}}\left[  2\left(  1-x_{\gamma}\right)
+a_{e}\right]  \frac{x_{\gamma}^{3}}{\left(  1-x_{\gamma}\right)  ^{2}%
}\nonumber\\
\mathcal{I}_{\gamma_{5}}\left[  P^{2},x_{\gamma}\right]   &  =\frac{-i}%
{\left(  4\pi\right)  ^{2}}\frac{1}{x_{\gamma}}\left(  F\left[  \frac{4m^{2}%
}{M^{2}}\right]  -F\left[  \frac{4m^{2}}{M^{2}}\frac{1}{1-x_{\gamma}}\right]
\right) \nonumber
\end{align}
where $m$ is the muon mass, $M$ the dimuonium mass, $x_{\gamma}$ the reduced
photon energy $2E_{\gamma}/M$, $a_{e}=4m_{e}^{2}/M^{2}$, $m_{e}$ the electron
mass, and with the function $F$ defined in (\ref{LoopFF}). In the limit
$x_{\gamma}\rightarrow0$, the spectrum $d\Gamma_{\gamma_{5}}/dx_{\gamma}$ goes
to zero as $x_{\gamma}^{3}$ as predicted by Low's theorem (the amplitude
behaves as $x_{\gamma}$, and an additional factor $x_{\gamma}$ comes from phase-space).

Taking the derivative of $\mathcal{I}_{\gamma_{5}}$ to get the corresponding
Coulomb form factor, we find
\begin{align*}
\mathcal{I}_{Coul}\left(  M^{2},x_{\gamma}\right)   &  =\left(  32\pi
C\phi_{o}\gamma\right)  \dfrac{\partial}{\partial M^{2}}\mathcal{I}%
_{\gamma_{5}}\left(  M^{2},x_{\gamma}\right) \\
&  =-i\frac{C\phi_{o}}{M}\frac{1}{x_{\gamma}}\left(  \frac{2}{\pi}%
\arctan\frac{M}{2\gamma}-\frac{4\gamma y_{\gamma}}{\pi M}\,\arctan y_{\gamma
}\right) \\
\text{where }\;y_{\gamma}  &  \equiv\left(  \frac{4m^{2}}{M^{2}\left(
1-x_{\gamma}\right)  }-1\right)  ^{-1/2}%
\end{align*}
The decay rate is then obtained by replacing $\mathcal{I}_{\gamma_{5}}$ by
$\mathcal{I}_{Coul}$ in (\ref{DalitzPhoton})%
\[
\Gamma\left(  p\text{-}Dm\rightarrow e^{+}e^{-}\gamma\right)  =\dfrac
{\alpha^{6}m}{6\pi}\frac{4m^{2}}{M^{2}}%
{\displaystyle\int\limits_{0}^{1-a_{e}}}
dx_{\gamma}\left|  \frac{2}{\pi}\arctan\frac{M}{2\gamma}-\frac{4\gamma
y_{\gamma}}{\pi M}\arctan y_{\gamma}\right|  ^{2}\frac{\rho\left(  x_{\gamma
},a_{e}\right)  }{x_{\gamma}^{2}}%
\]

It is very instructive to analyze in some details this result. Consider the
limit $\gamma\rightarrow0$ for the form factor:%
\begin{equation}
\underset{\gamma\rightarrow0}{\lim}\mathcal{I}_{Coul}\left(  M^{2},x_{\gamma
}\right)  =-i\frac{C\phi_{o}}{Mx_{\gamma}} \label{LimitG}%
\end{equation}
In that limit, the standard result for the decay rate is recovered%
\[
\Gamma\left(  p\text{-}Dm\rightarrow e^{+}e^{-}\gamma\right)  \overset
{\gamma\rightarrow0}{=}\dfrac{\alpha^{6}m}{6\pi}%
{\displaystyle\int_{0}^{1-a_{e}}}
dx_{\gamma}\frac{\rho\left(  x_{\gamma},a_{e}\right)  }{x_{\gamma}^{2}}%
\]
However, the differential rate $d\Gamma/dx_{\gamma}$has a wrong behavior when
$x_{\gamma}\rightarrow0$. The spectrum is linear (in $x_{\gamma}$) in the
limit $\gamma\rightarrow0$, in contradiction with Low's theorem. Therefore, it
appears that, contrary to the two real photon case, the limit $\gamma
\rightarrow0$ is far from smooth. It is inconsistent to consider both the
soft-photon limit and the on-shell limit simultaneously. Explicitly, the
incompatibility of the two limits is obvious if $x_{\gamma}\rightarrow0$ is
taken first
\begin{align}
\underset{x_{\gamma}\rightarrow0}{\lim}\mathcal{I}_{Coul}\left(
M^{2},x_{\gamma}\right)   &  =-i\frac{C\phi_{o}}{M\pi}\left(  \frac{M}%
{2\gamma}+\left(  1+\frac{M^{2}}{4\gamma^{2}}\right)  \arctan\frac{M}{2\gamma
}\right) \label{LimitX}\\
&  =-i\frac{C\phi_{o}}{M}\left(  \frac{M^{2}}{8\gamma^{2}}+\frac{1}%
{2}-\frac{4\gamma}{3M\pi}+...\right) \nonumber
\end{align}
Mathematically, what these considerations show is that the limit
$\gamma\rightarrow0$ does not exist at $x_{\gamma}=0$. Consequences of this
result were discussed in \cite{PreviousWork2}, but it should be clear that one
of the main virtue of our approach is its non-perturbative treatment of
$\gamma$, leading to correct photon spectra.

To close this section, we just mention that the second term of $\mathcal{I}%
_{Coul}\left(  M^{2},x_{\gamma}\right)  $ can be traced to the presence of
oblique cuts in the imaginary part (i.e. processes like $p$-$Dm\rightarrow
\mu^{+}\mu^{-}\gamma$ times $\mu^{+}\mu^{-}\left(  \gamma\right)  \rightarrow
e^{+}e^{-}\left(  \gamma\right)  $, see \cite{PreviousWork},
\cite{PreviousWork2}). Those are the processes neglected in standard
approaches, which is an inconsistent approximation since it is the sum of the
two terms of $\mathcal{I}_{Coul}\left(  M^{2},x_{\gamma}\right)  $ that
enforce Low's theorem. This can be seen by plotting
\[
\mathcal{J}\left(  x_{\gamma},\gamma\right)  =1-\frac{\frac{2\gamma}%
{M}y_{\gamma}\arctan y_{\gamma}}{\arctan\frac{M}{2\gamma}}%
\]
for $M=1$ and various values of $\gamma$%
\[%
{\includegraphics[
height=1.8645in,
width=3.0614in
]%
{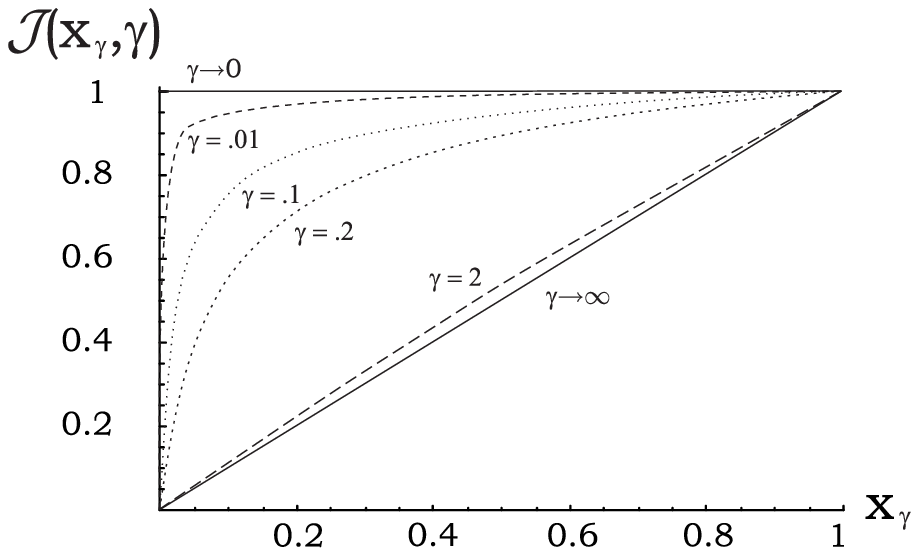}%
}%
\]
As the picture shows, even if $\gamma/M<<1$ for QED bound states, the limit
$\gamma\rightarrow0$ is not to be taken because it is singular.

\subsection{Orthodimuonium and Photon Vacuum Polarization}

We now consider the decay $o$-$Dm\rightarrow\gamma^{\ast}\rightarrow
e^{+}e^{-}$%
\[%
\raisebox{-0pt}{\includegraphics[
height=0.7965in,
width=2.2814in
]%
{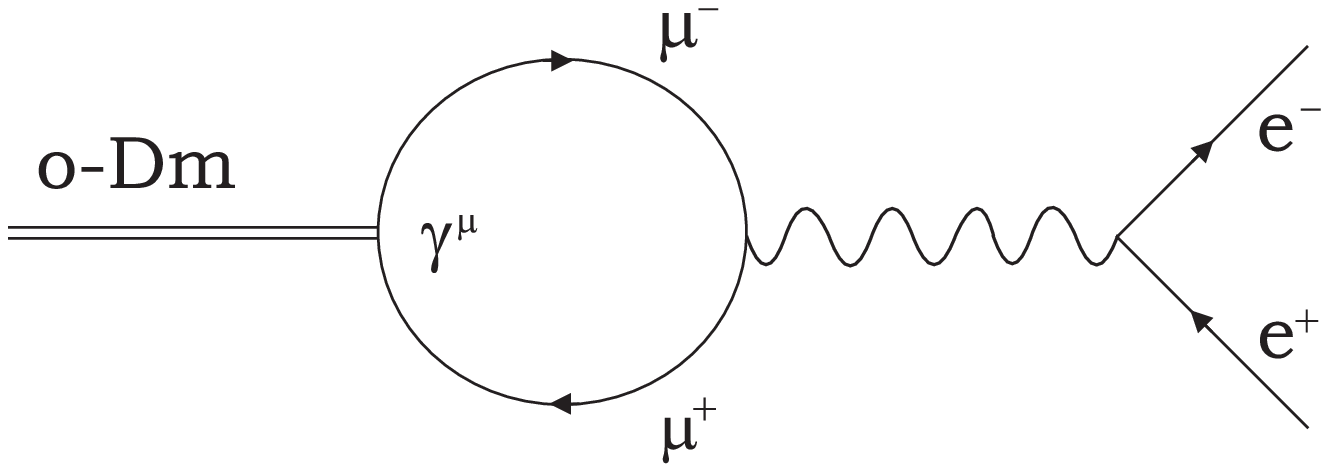}%
}%
\]
The point-like amplitude is easily obtained in terms of the (divergent) photon
vacuum polarization function
\[
\mathcal{M}_{\gamma^{\mu}}\left(  o\text{-}Dm\rightarrow e^{+}e^{-}\right)
=\varepsilon_{\mu}\left(  P\right)  \left(  P^{2}g^{\mu\nu}-P^{\mu}P^{\nu
}\right)  \Pi_{\gamma^{\mu}}\left(  P^{2}\right)  \frac{1}{P^{2}}\left\{
\overline{u}\left(  p\right)  \gamma_{\nu}v\left(  p^{\prime}\right)
\right\}
\]
where
\begin{equation}
\Pi_{\gamma^{\mu}}\left(  P^{2}\right)  =-\frac{e}{4\pi^{2}}\left[
\frac{D}{3}+\frac{5}{9}+\frac{4}{3\zeta}+\frac{2}{3}\left(  1-\frac{4}{\zeta
}\right)  \left(  1+\frac{2}{\zeta}\right)  \frac{\arctan\tfrac{1}%
{\sqrt{4/\zeta-1}}}{\sqrt{4/\zeta-1}}\right]  \label{PhotPol}%
\end{equation}
with $\zeta=P^{2}/m^{2}$, $m$ the muon mass and $D=2/\varepsilon
-\gamma_{Euler}+\log4\pi\mu^{2}/m^{2}$ in dimensional regularization.

The Coulomb form factor will be obtained from the derivative of the vacuum
polarization, with respect to $P^{2}=M^{2}$:
\begin{align}
\Pi_{Coul}\left(  M^{2}\right)   &  =\left(  32\pi C\phi_{o}\gamma\right)
\frac{\partial}{\partial M^{2}}\Pi_{\gamma^{\mu}}\left(  M^{2}\right)
\nonumber\\
&  =\frac{eC\phi_{o}}{M}\left[  \frac{8}{\pi}\left(  \gamma\frac{6m^{2}+M^{2}%
}{3M^{3}}-\frac{4m^{4}}{M^{4}}\arctan\frac{M}{2\gamma}\right)  \right]
\label{OrthoDalitz}\\
&  =\frac{eC\phi_{o}}{M}\left[  1-\frac{32}{3\pi}\frac{\gamma}{M}%
+8\frac{\gamma^{2}}{M^{2}}-\frac{128}{3\pi}\frac{\gamma^{3}}{M^{3}}+...\right]
\nonumber
\end{align}
This has the effect of removing the divergence, as it should. The decay rate
is
\begin{align*}
\Gamma\left(  o\text{-}Dm\rightarrow e^{+}e^{-}\right)   &  =\frac{\alpha}%
{3}\frac{M^{2}+2m_{e}^{2}}{M}\sqrt{1-a_{e}}\left|  \Pi_{Coul}\left(
M^{2}\right)  \right|  ^{2}\\
&  =\frac{\alpha^{5}m}{6}\left(  1+\frac{a_{e}}{2}\right)  \sqrt{1-a_{e}%
}\left|  1-\frac{32}{3\pi}\frac{\gamma}{M}+8\frac{\gamma^{2}}{M^{2}%
}-...\right|  ^{2}%
\end{align*}
To leading order, we recover the standard result $\Gamma\left(  o\text{-}%
Dm\rightarrow e^{+}e^{-}\right)  \approx\alpha^{5}m/6$ when $a_{e}<<1$. For
electromagnetic bound states, the binding energy is related to the fine
structure constant, hence the corrections can be cast into
\begin{align*}
\Gamma\left(  o\text{-}Dm\rightarrow e^{+}e^{-}\right)   &  =\frac{\alpha
^{5}m}{6}\left(  1+\frac{a_{e}}{2}\right)  \sqrt{1-a_{e}}\left(
1-\frac{16}{3}\frac{\alpha}{\pi}+\left(  \pi^{2}+\frac{64}{9}\right)
\frac{\alpha^{2}}{\pi^{2}}+...\right) \\
&  \approx\frac{\alpha^{5}m}{6}\left(  1+\frac{a_{e}}{2}\right)  \sqrt
{1-a_{e}}\left(  1-1.70\alpha+1.72\alpha^{2}+...\right)
\end{align*}
Compared to the binding energy corrections to the parapositronium two-photon
decay rate, the present corrections are much bigger. Note also that the
binding energy correction obtained here, at order $\alpha$, again account for
a great deal of the correction obtained using the standard approach (we
consider only part of the total $\mathcal{O}\left(  \alpha\right)  $
correction, see Eq. 47 in \cite{Dimuonium}), which are
\[
\Gamma^{LO+NLO}\left(  o\text{-}Dm\rightarrow e^{+}e^{-}\right)
=\frac{\alpha^{5}m}{6}\left(  1+\frac{a_{e}}{2}\right)  \sqrt{1-a_{e}}\left(
1-4\frac{\alpha}{\pi}+...\right)
\]

\section{Orthopositronium Decay}

We now apply our method to the orthopositronium decay to three photons. This
is a very interesting decay process. As discussed in \cite{PreviousWork2}, the
lowest order basis chosen in standard computations, namely the Ore-Powell
amplitude \cite{OrePowell}, is in contradiction with Low's theorem. What we
will now show is that taking binding energy into account, i.e. integrating the
effects of all the Coulomb photon exchanges at lowest order through the
appropriate form factor, is necessary in order to get a correct spectrum.
Also, we will again find that the bulk of the radiative corrections is
accounted for already at our lowest order.

\subsection{Point-like Amplitude}

The point-like amplitude is the standard light-by-light box diagram%
\[%
{\includegraphics[
height=1.2756in,
width=1.8697in
]%
{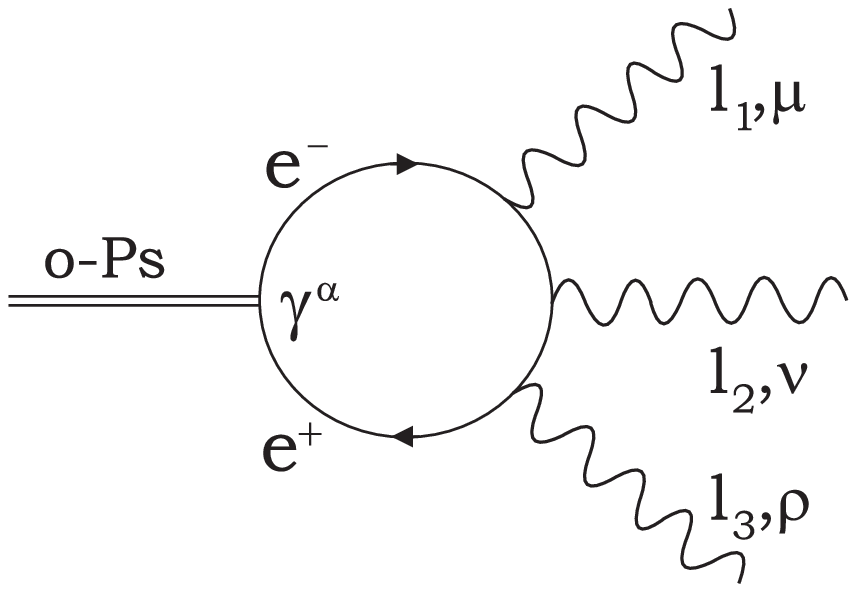}%
}%
\]
plus five other ordering of the photon insertions. The amplitude can be found
in many places \cite{Anton}, \cite{LightbyLight}. The tensor $G_{\alpha
}^{\lambda_{1}\lambda_{2}\lambda_{3}}$ describing the transition from an
off-shell photon to three on-shell photons of helicity states $\lambda
_{1}\lambda_{2}\lambda_{3}$ is such that
\begin{equation}
\sum_{\lambda_{1}\lambda_{2}\lambda_{3}}\left(  G_{\alpha}^{\lambda_{1}%
\lambda_{2}\lambda_{3}}G^{\ast\lambda_{1}\lambda_{2}\lambda_{3},\alpha
}\right)  =\frac{2^{4}\alpha^{3}}{\pi}\left[  R\left(  123\right)  +R\left(
213\right)  +R\left(  312\right)  \right]  \label{ThreContr}%
\end{equation}
where $R\left(  123\right)  \equiv R\left(  x_{1},x_{2},x_{3},a\right)  $
($x_{i} $ is the reduced energy of the photon $i$ and $a=4m^{2}/M^{2}$). The
$R\left(  ijk\right)  $ are given in terms of individual dimensionless
helicity amplitudes as
\begin{align}
R\left(  123\right)   &  =\frac{1}{3}\left|  E_{-++}^{\left(  2\right)
}\left(  123\right)  \right|  ^{2}+\left|  E_{+++}^{\left(  2\right)  }\left(
123\right)  \right|  ^{2}\label{RR}\\
&  +\frac{x_{1}}{x_{2}x_{3}\left(  1-x_{1}\right)  }\left|  E_{-++}^{\left(
1\right)  }\left(  213\right)  \right|  ^{2}+\frac{1}{x_{1}^{2}}\left|
E_{+++}^{\left(  1\right)  }\left(  123\right)  +E_{+++}^{\left(  1\right)
}\left(  132\right)  \right|  ^{2}\nonumber\\
&  +\frac{\left(  1-x_{2}\right)  \left(  1-x_{3}\right)  }{x_{1}^{2}\left(
1-x_{1}\right)  }\left|  \frac{1}{1-x_{2}}E_{+++}^{\left(  1\right)  }\left(
123\right)  -\frac{1}{1-x_{3}}E_{+++}^{\left(  1\right)  }\left(  132\right)
\right|  ^{2}\nonumber
\end{align}
The helicity amplitudes $E_{\pm++}^{\left(  n\right)  }$ are complicated
functions of $x_{i}$ and $a$, and we do not reproduce them here (\cite{Anton},
\cite{LightbyLight}). The decay rate of a point-like vector positronium to
three photons is then%
\[
\Gamma_{\gamma^{\mu}}\left(  o\text{-}Ps\rightarrow\gamma\gamma\gamma\right)
=\frac{1}{9}\frac{\alpha^{3}}{2^{5}\pi^{4}}M\int dx_{1}dx_{2}\left[  R\left(
123\right)  +R\left(  213\right)  +R\left(  312\right)  \right]
\]
with the three-body phase space written in terms of reduced photon energies as
$\int d\Phi_{3}=M^{2}\int dx_{1}dx_{2}/2^{7}\pi^{3}$ \cite{Anton}.

Of special interest is the behavior of the low-energy end of the differential
rate. For various values of the ratio $a=4m^{2}/M^{2}$ (i.e. of the binding
energy $E_{B}=M-2m)$, the photon spectrum, normalized to the total rate is%
\[%
{\includegraphics[
height=2.1369in,
width=3.103in
]%
{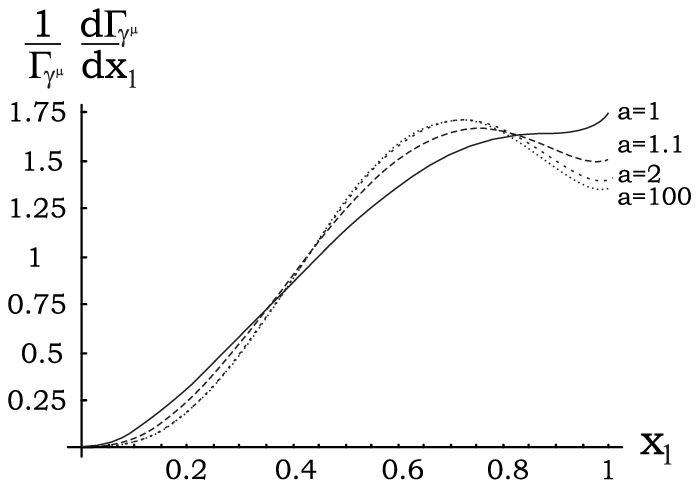}%
}%
\]
Sufficiently close to zero, the behavior is always like $x_{1}^{3}$, as
required by Low's theorem \cite{PreviousWork2}. As $a$ increases, the $x^{3}$
behavior is getting more and more pronounced. In the limit $a\rightarrow
\infty$, the normalized spectrum is%
\[
\frac{1}{\Gamma_{\gamma^{\mu}}}\frac{d\Gamma_{\gamma^{\mu}}}{dx_{1}}%
\overset{a\rightarrow\infty}{=}\frac{5}{17}x_{1}^{3}\left(  \frac{343}%
{3}-207x_{1}+\frac{973}{10}x_{1}^{2}\right)
\]
which is, as expected, the spectrum obtained from the Euler-Heisenberg
effective theory \cite{EH}
\[
\mathfrak{L}_{E-H}=\frac{\alpha^{2}}{90m^{4}}\left[  \left(  F_{\mu\nu}%
F^{\mu\nu}\right)  ^{2}+\frac{7}{4}(F_{\mu\nu}\widetilde{F}^{\mu\nu}%
)^{2}\right]
\]

\subsection{Coulomb Form factor and Ore-Powell spectrum}

To insert the Schr\"{o}dinger wavefunction form factor, we define modified
helicity amplitudes according to (\ref{Derivative2})%
\[
E_{\pm++,Coul}^{\left(  n\right)  }\left(  ijk,M^{2}\right)  =\left(  32\pi
C\phi_{o}\gamma\right)  \dfrac{\partial}{\partial M^{2}}E_{\pm++}^{\left(
n\right)  }\left(  ijk,M^{2}\right)
\]
The resulting decay amplitude, and decay rate are constructed as in the
punctual case, and we reach
\[%
\begin{array}
[c]{l}%
\Gamma\left(  o\text{-}Ps\rightarrow\gamma\gamma\gamma\right)  =\alpha
^{6}m\dfrac{2}{9\pi}\left(  \dfrac{4m^{2}}{M^{2}}\right)  ^{2}\left(
\dfrac{\mathcal{R}}{2\pi^{2}}\right) \\
\text{with }\mathcal{R}\equiv%
{\displaystyle\int}
dx_{1}dx_{2}\left(  R_{Coul}\left(  123\right)  +R_{Coul}\left(  213\right)
+R_{Coul}\left(  312\right)  \right)
\end{array}
\]
$R_{Coul}$ is given by (\ref{RR}) with $E_{\pm++,Coul}^{\left(  i\right)  }$
in place of $E_{\pm++}^{\left(  i\right)  }$. Integrating over $x_{2}$, we get
the photon spectrum for various values of $a$
\begin{equation}%
{\includegraphics[
height=2.7095in,
width=4.0162in
]%
{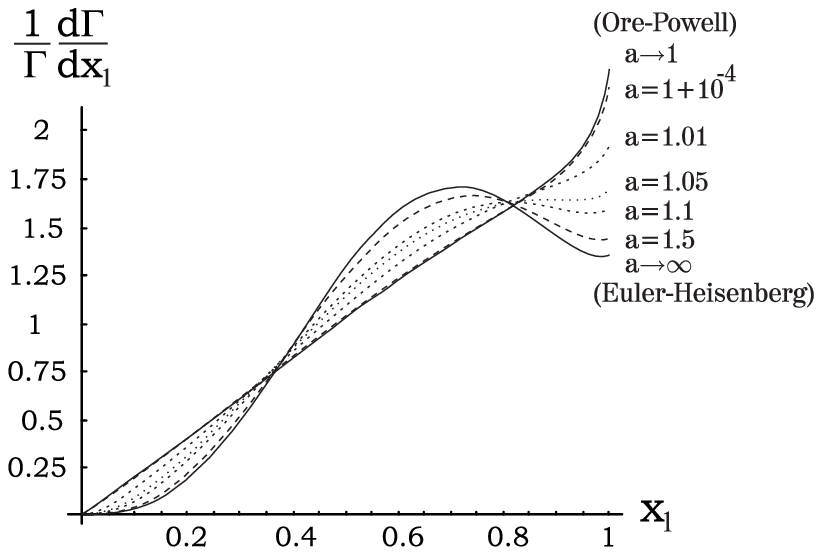}%
}%
\label{FigOre}%
\end{equation}

As long as $a\neq1$, the spectrum behavior is in $x_{1}^{3}$ close to zero,
i.e. roughly in the range $x_{1}\in\left[  0,a-1\right]  $. The properties of
this spectrum as $a$ varies are completely similar to that of $p$%
-$Dm\rightarrow\gamma e^{+}e^{-}$, and one can show that the limits
$\gamma\rightarrow0$ and $x_{1}\rightarrow0$ are again incompatible.

In the context of dispersion relations, the contributions originating in the
oblique cuts (processes like $o$-$Ps\rightarrow e^{+}e^{-}\gamma$ times
$e^{+}e^{-}\left(  \gamma\right)  \rightarrow\gamma\gamma\left(
\gamma\right)  $ in the imaginary part) are essential to maintain a physical
spectrum (i.e. in agreement with Low's theorem). Again, those oblique cuts are
neglected in standard approaches. Here, they are automatically accounted for
since they are included in the point-like result.

Note also that the spectrum for a point-like bound state at threshold
corresponds to the spectrum for $a\sim1.05$ in the Coulomb form factor case.
The bound state decay spectra for $a<1.05$ are unattainable in the point bound
state case.

Concerning the total integrated rate, we recover the Ore-Powell result at
threshold $\left(  a=1\right)  $%
\[
\Gamma\left(  o\text{-}Ps\rightarrow\gamma\gamma\gamma\right)  =\alpha
^{6}m\frac{2}{9\pi}\left(  \frac{17.16}{2\pi^{2}}\right)  =4.75\times
10^{-15}\;\text{MeV }=\alpha^{6}m\frac{2}{9\pi}\left(  \pi^{2}-9\right)
\]
As $a$ increases, the total rate quickly decreases (remember $a=4\gamma
^{2}/M^{2}+1$)%
\[%
{\includegraphics[
height=1.4598in,
width=2.1395in
]%
{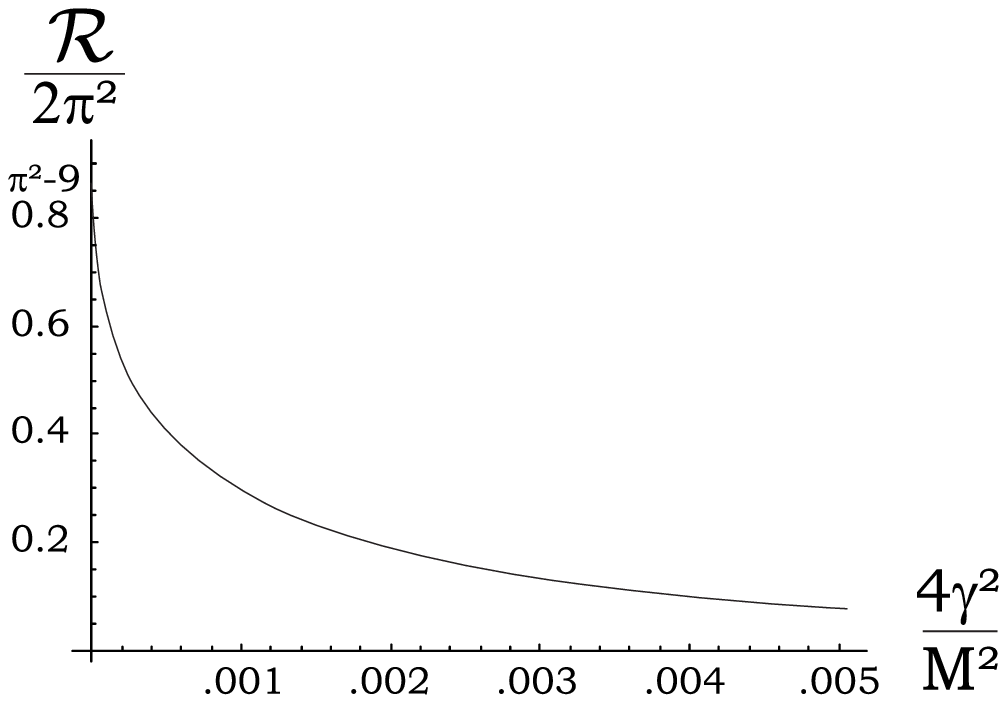}%
}%
\]
By a numerical analysis, we found
\[
\frac{\mathcal{R}}{2\pi^{2}}=\left(  \pi^{2}-9\right)  \left(
1-15.412\frac{\gamma}{M}+122\frac{\gamma^{2}}{M^{2}}-889\frac{\gamma^{3}%
}{M^{3}}+1.92\;10^{4}\;\frac{\gamma^{4}}{M^{4}}-...\right)
\]
For orthopositronium, we can express $\gamma$ in terms of $\alpha$, and we
find the binding energy corrections to the total rate
\begin{equation}
\Gamma\left(  o\text{-}Ps\rightarrow\gamma\gamma\gamma\right)  =\alpha
^{6}m\frac{2\left(  \pi^{2}-9\right)  }{9\pi}\left(  1-12.1\frac{\alpha}{\pi
}+80.2\frac{\alpha^{2}}{\pi^{2}}-502\frac{\alpha^{3}}{\pi^{3}}+...\right)
\label{OrthoGam}%
\end{equation}
This series is to be compared to the one presented in the literature
\cite{RadiativeCorr}, which is
\begin{subequations}
\begin{align*}
\Gamma_{o\text{-}Ps}  &  =\alpha^{6}m\frac{2\left(  \pi^{2}-9\right)  }{9\pi
}\left(  1-A_{o}\frac{\alpha}{\pi}-\frac{\alpha^{2}}{3}\ln\frac{1}{\alpha
}+B_{o}\frac{\alpha^{2}}{\pi^{2}}-\frac{3\alpha^{3}}{2\pi}\ln^{2}%
\frac{1}{\alpha}\right. \\
&  \;\;\;\;\;\;\;\;\;\;\;\;\;\;\;\;\;\;\;\;\;\;\;\;\;\;\;\;\;\;\;\left.
+C_{o}\frac{\alpha^{3}}{\pi}\ln\frac{1}{\alpha}+\delta_{5\gamma}%
\frac{\alpha^{2}}{\pi^{2}}\right)
\end{align*}
with coefficients
\end{subequations}
\[%
\begin{tabular}
[c]{l}%
$A_{o}=10.286606\left(  10\right)  $\\
$B_{o}=44.52\left(  26\right)  $%
\end{tabular}
\;\;\;\;%
\begin{tabular}
[c]{l}%
$C_{o}=5.517\left(  1\right)  $\\
$\delta_{5\gamma}=0.19\left(  1\right)  $%
\end{tabular}
\]
Again, the exchanges of Coulomb photon appear responsible for the bulk of the
radiative corrections at order $\alpha$. Note that the origin of the slowness
in the convergence of the corrections to $\Gamma_{o\text{-}Ps}$ is clearly
identified as coming from the perturbative expansion of binding energy effects
(see (\ref{OrthoGam})). In other words, the four-point fermionic loop with a
Coulomb form factor is not well-behaved for $\gamma=0$, and that limit appears
as an inappropriate basis for perturbation theory.

\section{Application to other processes}

In this final section, we present two extensions of the method. First, we
accommodate the derivative formula (\ref{Derivative2}) for radial excitations,
by considering the pion decay $\pi^{0}\rightarrow\gamma\,o$-$Ps$. For the
second, we compute the hyperfine splitting using the language of mass renormalization.

\subsection{Pion Decay to Orthopositronium \newline and Radial Excitations}

The technique of taking the derivative of point-like amplitude can be extended
to other spherically symmetric wavefunctions. The decay $\pi^{0}%
\rightarrow\gamma\,o$-$Ps$ is a good example \cite{Pion}%
\[%
{\includegraphics[
height=1.0784in,
width=1.6137in
]%
{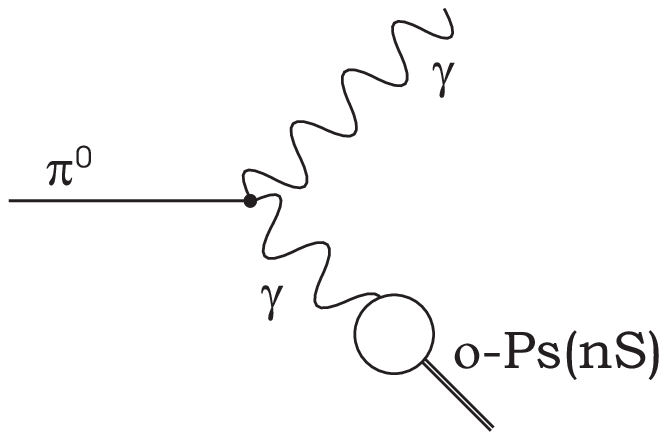}%
}%
\]
To get a sensible theoretical prediction, one must sum the decay rates over
the infinite tower of radial excitations $o$-$Ps\left(  nS\right)  $ in the
final state ($o$-$Ps\left(  nS\right)  $ means $n$th radial excitation of the
$S$-wave $J=1$ positronium state). From the standard Schr\"{o}dinger
wavefunction for hydrogen states, we can write a general expression for the
form factor for $nS$ radial excitations as%
\begin{equation}
\mathcal{I}_{Coul,n}\left(  M_{n}^{2}\right)  =\left(  32\pi C\phi
_{noo}\right)  \,\left[  _{1}F_{1}\left(  1-n,2,16\gamma_{n}^{2}%
\dfrac{\partial}{\partial M_{n}^{2}}\right)  \right]  \left[  \gamma
_{n}\left(  M_{n}^{2}\right)  \dfrac{\partial}{\partial M_{n}^{2}}%
\mathcal{I}_{\gamma^{\mu}}\left(  M_{n}^{2}\right)  \right]  \label{Radial}%
\end{equation}
with%
\[
\left|  \phi_{noo}\right|  =\sqrt{\dfrac{m^{3}\alpha^{3}}{8\pi n^{3}}%
}\,\;\;\gamma_{n}=\sqrt{m^{2}-M_{n}^{2}/4}%
\]
Where we have denoted $M_{n}$ the mass of the $o$-$Ps\left(  nS\right)  $
state. The hypergeometric functions are essentially the well-known Laguerre
polynomials ((\ref{Radial}) is equally valid for parapositronium).

For the case at hand, $\mathcal{I}_{\gamma^{\mu}}$ is the photon vacuum
polarization (\ref{PhotPol}), where $m$ refers now to the electron mass. The
pion decay rate into orthopositronium states can be written as%
\[
R_{oPs\left(  nS\right)  }=\frac{\Gamma\left(  \pi^{0}\rightarrow
\gamma\,o\text{-}Ps\left(  nS\right)  \right)  }{\Gamma\left(  \pi
^{0}\rightarrow\gamma\gamma\right)  }=2e^{2}\left|  \Pi_{Coul,n}\left(
M_{n}^{2}\right)  \right|  ^{2}\left(  1-\frac{M_{n}^{2}}{m_{\pi}^{2}}\right)
^{3}\left[  1+\mathcal{O}\left(  \frac{M_{n}^{2}}{m_{\rho,\omega}^{2}}\right)
\right]
\]
The $M_{n}^{2}/m_{\rho,\omega}^{2}\approx10^{-6}$ corrections arise from the
form factor for pion to two photons. The mass ratio $M_{n}^{2}/m_{\pi}%
^{2}\approx10^{-4}$ is also negligible compared to binding energy corrections,
to which we now turn. Up to corrections of order $\gamma_{n}^{2}$, we can
write using $\gamma_{n}\approx m\alpha/2n$
\[
R_{oPs\left(  nS\right)  }=\dfrac{\alpha^{4}}{2}\frac{1}{n^{3}}\left(
1-2A_{n}\frac{\gamma_{n}}{M_{n}}+...\right)  =\dfrac{\alpha^{4}}{2}%
\frac{1}{n^{3}}\left(  1-\frac{1}{2n}A_{n}\alpha+...\right)
\]
(the $A_{n}$ are the numerical coefficients found by expanding $\Pi
_{Coul,n}\left(  M_{n}^{2}\right)  $). Summing over $n$, we find%
\[
\sum_{n}R_{oPs\left(  nS\right)  }^{LO}=\dfrac{\alpha^{4}}{2}\zeta\left(
3\right)  \left(  1-1.66\alpha\right)  \approx1.684\times10^{-9}%
\]
where $\alpha^{4}/2\approx\left(  1.418\times10^{-9}\right)  $ is obtained
from the contribution of the $o$-$Ps\left(  1S\right)  $ only. For comparison,
\cite{Pion} found the radiative correction to be $\left(  1-0.92\alpha\right)
$. The experimentally quoted branching fraction is \cite{Pion2}%
\[
\left.  \frac{\Gamma\left(  \pi\rightarrow\gamma Ps\right)  }{\Gamma\left(
\pi^{0}\rightarrow\gamma\gamma\right)  }\right|  _{\exp}=\left(
1.9\pm0.3\right)  \times10^{-9}%
\]
so the agreement is good.

What this little exercise shows is the power of our method as a mean to
partially compute higher order corrections. More importantly, it is by now
apparent that other, more complicated wavefunctions can easily be accommodated
for. Any wavefunction that can be expressed, even approximately, from
derivatives of the Coulomb one can fit in our scheme. This may open the way to
many applications in QCD, as we will comment in the conclusion.

\subsection{Hyperfine Splitting \newline and Mass Renormalization}

We implement here the renormalization of the positronium state. The idea is to
consider the bare positronium mass $M$ as equal to twice the electron mass
$m$, and then to carry a (finite) mass renormalization. The diagrams
contributing, at lowest order, to this mass shift will be obtained as the
second derivatives of the vacuum polarization loops:
\[%
{\includegraphics[
height=0.6832in,
width=3.3486in
]%
{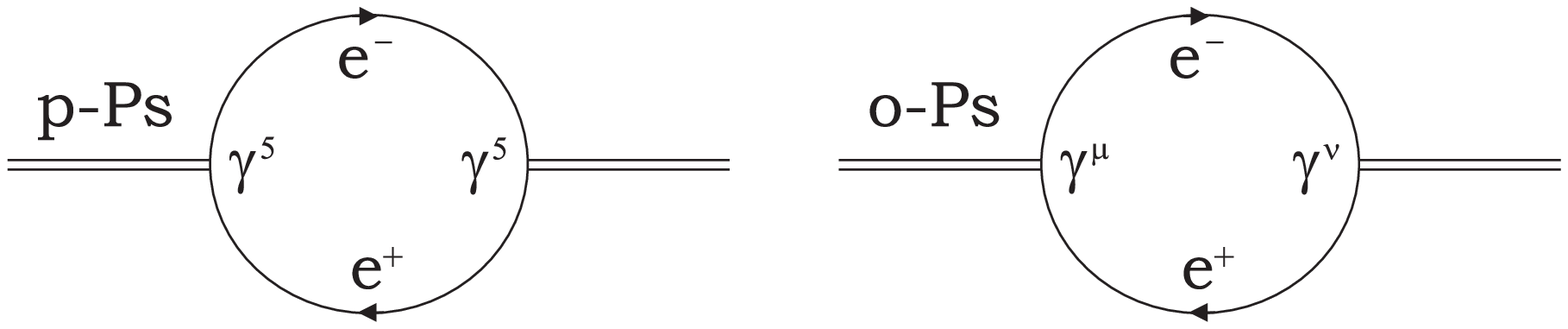}%
}%
\]
Let us first obtain the mass renormalization equations, and then discuss the
results numerically.

\subsubsection{Mass Renormalization}

For a pseudoscalar parapositronium state, the resummation of the Dyson series
is trivial. The parapositronium propagator is then
\[
G\left(  q^{2}\right)  =\frac{i}{q^{2}-M_{o}^{2}-\Pi_{para}\left(
q^{2}\right)  }%
\]
Of interest to us is the mass shift, defined from the pole of $G\left(
q^{2}\right)  $
\begin{equation}
M_{R,para}^{2}-M_{o}^{2}=\Pi_{para}\left(  M_{R}^{2}\right)  \rightarrow
M_{R,para}-M_{o}=\frac{\Pi_{para}\left(  M_{R,para}^{2}\right)  }{2M_{o}}
\label{ParaRen}%
\end{equation}

For the orthopositronium, the transverse part of the bare propagator is
\[
G_{0}^{\mu\nu}\left(  q^{2}\right)  =\frac{-i\left(  g^{\mu\nu}-\frac{q^{\mu
}q^{\nu}}{q^{2}}\right)  }{q^{2}-M_{0}^{2}+i\varepsilon}%
\]
The self-energy of the vector positronium will be of the form
\[
\Pi^{\mu\nu}\left(  q^{2}\right)  =\left(  q^{2}g^{\mu\nu}-q^{\mu}q^{\nu
}\right)  \Pi_{ortho}\left(  q^{2}\right)
\]
Proceeding with the resummation of the Dyson series, we end up with
\[
G^{\mu\nu}\left(  q^{2}\right)  =\frac{-i\left(  g^{\mu\nu}-\frac{q^{\mu
}q^{\nu}}{q^{2}}\right)  }{q^{2}\left(  1-\Pi_{ortho}\left(  q^{2}\right)
\right)  -M_{o}^{2}}%
\]
Again, the propagator pole is at $q^{2}=M_{R}^{2}$, hence
\begin{equation}
M_{R,ortho}^{2}-M_{o}^{2}=M_{R}^{2}\Pi_{ortho}\left(  M_{R}^{2}\right)
\rightarrow M_{R,ortho}-M_{o}=\frac{M_{R,ortho}}{2}\Pi_{ortho}\left(
M_{R,ortho}^{2}\right)  \label{OrthoRen}%
\end{equation}

The two equations (\ref{ParaRen}) and (\ref{OrthoRen}) are self-consistent
equations. By identifying $M_{o}=2m$ and $M_{R,para\left(  ortho\right)
}=M_{para\left(  ortho\right)  }$, they are of the form%
\[
\left\{
\begin{array}
[c]{l}%
E_{B}^{para}\equiv M_{para}-2m=\dfrac{\Pi_{para}\left(  M_{para}^{2}\right)
}{4m}\equiv f^{para}\left(  E_{B}^{para}\right) \\
E_{B}^{ortho}\equiv M_{ortho}-2m=\dfrac{M_{ortho}}{2}\Pi_{ortho}\left(
M_{ortho}^{2}\right)  \equiv f^{ortho}\left(  E_{B}^{ortho}\right)
\end{array}
\right.
\]
Our approach will be to generate corrections to $E_{B}^{para\left(
ortho\right)  }$ by plugging the non-relativistic result $E_{B}=-m\alpha
^{2}/4$ into $f^{para\left(  ortho\right)  }$.

\subsubsection{Second Derivatives and Hyperfine Splitting}

The double derivatives of the vacuum polarization loops are%
\[
\Pi_{para\left(  ortho\right)  }\left(  M^{2}\right)  =\frac{1}{2}\left(
32\pi C\phi_{o}\gamma\right)  ^{2}\frac{1}{2}\left(  \dfrac{\partial}{\partial
M^{2}}\right)  ^{2}\Pi_{\gamma^{5}\left(  \gamma^{\mu}\right)  }\left(
M^{2}\right)
\]
where a factor of $1/2$ corrects for the factor $2$ appearing in the
derivative of $1/\left(  s-M^{2}\right)  ^{2}$, while the other accounts for
the double counting of Coulomb photon exchanges.

The computations are straightforward. For parapositronium, the point-like
quadratic divergence disappears and we find
\begin{align*}
E_{B}^{para}  &  \equiv M_{para}-2m=\frac{\Pi_{para}\left(  M_{para}%
^{2}\right)  }{4m}\\
&  =-\frac{4\left|  \phi_{o}\right|  ^{2}}{Mm}\left[  \left(  \frac{4\gamma
^{2}}{M^{2}}+1\right)  \frac{\arctan\frac{M}{2\gamma}}{2\gamma/M}%
-\frac{\frac{4\gamma^{2}}{M^{2}}-1}{\frac{4\gamma^{2}}{M^{2}}+1}\right]
_{M=M_{para}}\\
&  =-\frac{m\alpha^{2}}{4}-\frac{9m\alpha^{4}}{128}+\frac{m\alpha^{5}}{6\pi
}-\frac{115m\alpha^{6}}{8192}+...
\end{align*}
Where we have used $\left|  \phi_{o}\right|  ^{2}=m^{3}\alpha^{3}/8\pi
$,$\gamma^{2}=m^{2}-M_{para}^{2}/4$ and $M_{para}=2m-m\alpha^{2}/4$ in the
right hand side. Proceeding similarly with the orthopositronium
renormalization, we consider the second derivative of the photon vacuum
polarization function:%
\begin{align}
E_{B\left(  1\right)  }^{ortho}  &  \equiv M_{ortho}-2m=\frac{M_{ortho}}{2}%
\Pi_{ortho}\left(  M_{ortho}^{2}\right) \label{OrthoBind}\\
&  =\frac{4\left|  \phi_{o}\right|  ^{2}}{M^{2}}\left[  \left(  \tfrac
{4\gamma^{2}}{M^{2}}+1\right)  \left(  \tfrac{20\gamma^{2}}{M^{2}}-1\right)
\frac{\arctan\frac{M}{2\gamma}}{\gamma/M}-\tfrac{2+\frac{176}{3}%
\frac{\gamma^{2}}{M^{2}}+160\frac{\gamma^{4}}{M^{4}}}{\frac{4\gamma^{2}}%
{M^{2}}+1}\right]  _{M=M_{ortho}}\nonumber\\
&  =-\frac{m\alpha^{2}}{4}+\frac{27m\alpha^{4}}{128}-\frac{2m\alpha^{5}}{3\pi
}+\frac{1301m\alpha^{6}}{8192}+...\nonumber
\end{align}
At the order $\alpha^{4}$, there is also the annihilation diagram
\[%
{\includegraphics[
height=0.5967in,
width=3.0268in
]%
{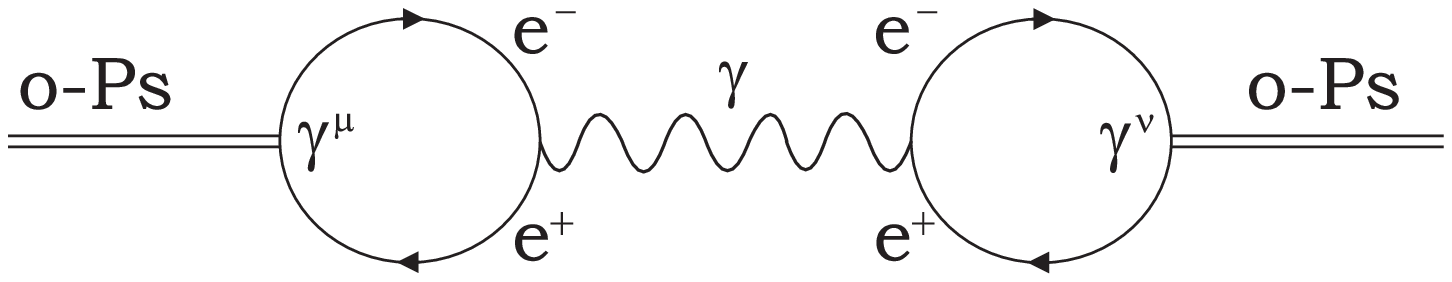}%
}%
\]
whose contribution is $\Pi_{ortho}^{Ann,\mu\nu}\left(  M^{2}\right)
=-i\left(  P^{2}g^{\mu\nu}-P^{\mu}P^{\nu}\right)  \left(  \Pi_{Coul}\left(
P^{2}\right)  \right)  ^{2}$. Using $\Pi_{Coul}\left(  P^{2}\right)  $ as
given by (\ref{OrthoDalitz}), the dominant contribution in the limit
$\gamma\rightarrow0$ is%
\[
\Pi_{ortho}^{Ann}\left(  P^{2}\right)  =\frac{\alpha^{4}}{4}%
+...\;\;\rightarrow\;\;E_{B\left(  2\right)  }^{ortho,Ann}=\frac{M_{ortho}}%
{2}\Pi_{ortho}^{Ann}\left(  M_{ortho}^{2}\right)  =\frac{m\alpha^{4}}%
{4}+...\text{ }%
\]
Hence the lowest loop contributions to the hyperfine splitting are, to order
$\alpha^{4}$:%
\[
\Delta E_{hf}=\left(  E_{B\left(  1\right)  }^{ortho}+E_{B\left(  2\right)
}^{ortho,Ann}\right)  -E_{B}^{para}=m\alpha^{4}\left[  \frac{17}%
{32}+...\right]  =m\alpha^{4}\left[  0.5313+...\right]
\]

This is to be compared to the result (see \cite{QEDTest}, \cite{HF} and
references cited there)
\begin{align*}
\Delta E_{hf}  &  =m\alpha^{4}\left\{  \frac{7}{12}-\frac{\alpha}{\pi}\left(
\frac{8}{9}+\frac{\ln2}{2}\right)  \right. \\
&  +\frac{\alpha^{2}}{\pi^{2}}\left(  -\frac{5}{24}\pi^{2}\ln\alpha
+\frac{1367}{648}-\frac{5197}{3456}\pi^{2}+\left(  \frac{221}{144}\pi
^{2}+\frac{1}{2}\right)  \ln2-\frac{53}{32}\zeta\left(  3\right)  \right) \\
&  \left.  +\frac{\alpha^{3}}{\pi^{3}}\left(  -\frac{7}{8}\pi^{2}\ln^{2}%
\alpha+\left(  \frac{17}{3}\ln2-\frac{217}{90}\right)  \pi^{3}\ln
\alpha\right)  +\mathcal{O}\left(  \alpha^{3}\right)  \right\} \\
&  \approx m\alpha^{4}\left[  0.5833-0.3933\alpha-0.2083\alpha^{2}\ln
\alpha-0.3928\alpha^{2}-...\right]
\end{align*}
where the correction of order $\mathcal{O}\left(  \alpha^{7}\ln\alpha\right)
$ has been obtained very recently \cite{HF}.

It is not surprising that our method does not reproduce exactly the above
result, because we have neglected many diagrams (like electron self-energy
insertions for example), and because we used the lowest order non-relativistic
binding energy $E_{B}=-m\alpha^{2}/4$ and wavefunction $\left|  \phi
_{o}\right|  ^{2}=m^{3}\alpha^{3}/8\pi$ as a basis. Taken individually, the
corrections of $\mathcal{O}\left(  \alpha^{4}\right)  $ to the para- and
orthopositronium masses are off by more than $50\%$. On the other hand, the
difference between both corrections, giving the hyperfine splitting, is
surprisingly good. Again, it seems that all the effects contained in the
lowest order loop (i.e. all the ladder Coulomb photon exchanges) suffice to
account for most of the radiative corrections. In conclusion, further studies
of the application of our method to hyperfine splitting appear as necessary.

\section{Conclusion}

We can characterize our method as a change of basis for perturbation theory:
some binding effects are treated at all orders in $\gamma$, at each order in
$\alpha$. With such non-perturbative results, the behavior of the usual
expansion in the binding energy can be analyzed. \textit{We find that such
expansions are problematic when soft-photons are present (}i.e. when a
photon's energy is of the order of $\gamma$). In other words, expansions in
the binding energy are incompatible with analyticity. In all cases, unphysical
contributions violating analyticity are subleading, being of the order of
$2m-M\approx m\alpha^{2}/4$. However small, such effects become important in
view of the precision achieved in both experimental and theoretical
descriptions. In addition, standard positronium models neglect some
contributions (corresponding, in the context of dispersion relations, to
oblique cuts contributions to the imaginary part of the loop diagrams).
Therefore, they cannot be correct at order $\alpha^{2}$ for decay processes
\cite{PreviousWork}.

In our view, obtaining QED bound state amplitudes from relativistic point-like
amplitudes has many advantages. This may lead to simplifications in the
computation of higher order corrections, especially concerning issues of gauge
invariance and infrared divergence. For instance, two-loop graphs will
introduce some $O\left(  \alpha^{2}\right)  $ and $O\left(  \alpha^{2}%
\ln\alpha\right)  $ corrections (see for example \cite{VacPolPole}). We should
point-out, however, that the technique of taking the derivative will not work
as it stands for higher order diagrams, because of binding graphs, and some
refinements will be necessary.

As a final comment, the present method can lead to interesting advances in
quarkonium physics. Of course, some knowledge of the quarkonium wavefunctions
will be necessary. No matter the form of these wavefunctions, it should be
possible to expand them in the basis made of the Coulomb wavefunction and its
derivatives, to which the present method apply. This is left to further
studies. For now, as a first result, it appear quite obvious that the
Ore-Powell spectrum used as a lowest order approximation for the photon
spectrum in the quarkonium inclusive decays into hadrons + photon is
incorrect. If quarkonia were Coulomb bound state, the spectrum would be the
curve of (\ref{FigOre}) with $a\approx1.2\rightarrow1.5$ (since $4m_{B}%
^{2}/M_{\Upsilon\left(  1S\right)  }^{2}\approx1.25$ and $4m_{D}^{2}%
/M_{J/\psi}^{2}\approx1.46$), i.e. a spectrum which is suppressed at high
energy compared to the Ore-Powell one. QCD effects should not change much our
conclusion. We think that the suppression at high energy observed by CLEO
\cite{PromptPhoton} is, at least in part, a manifestation of binding energy
effects, giving us confidence in our results.

\ \newline 

{\Large Acknowledgments: }We are very pleased to acknowledge useful
discussions with Gabriel Lopez Castro and St\'{e}phanie Trine. C. S.
acknowledges financial support from IISN (Belgium).

\end{document}